# Electronic properties of edge-functionalized zigzag graphene nanoribbons on SiO$_2$ substrate

D. M. Zhang, Z. Li, J. F. Zhong, L. Miao†, J. J. Jiang

*Department of Electronic Science and Technology, Huazhong University of Science and Technology, Wuhan, HUBEI 430074, People's Republic of China*

Based on first-principles calculations, electronic properties of edge-functionalized zigzag graphene nanoribbons (ZGNRs) on SiO$_2$ substrate are presented. Metallic or semiconducting properties of ZGNRs are revealed due to various interactions between edge-hydrogenated ZGNRs and different SiO$_2$ (0001) surfaces. Bivalent functional groups decorating ZGNRs serve as the bridge between active edges of ZGNRs and SiO$_2$. These functional groups stabilize ZGNRs on substrate, as well as modify the edge states of ZGNRs and further affect their electronic properties. Band gaps are opened owing to edge states destruction and distorted lattice in ZGNRs.

PACS number(s): 73.22.Pr, 73.40.Ty, 71.15.Mb

## I. INTRODUCTION

Graphene, an atomic monolayer of bulk graphite, has potential applications in nanoelectronics due to its unique electronic properties.[1-3] However, two major difficulties remain in graphene-based devices，manufacture of large-scale and high-quality graphene, as well as controlling of the electronic structures. Recently, various experimental and theoretical researches about substrates, which support large-scale and high-quality graphene sheets in addition to opening a gap in graphene, seemingly have tackled the problems in graphene-base nano devices. Fine graphene sheets on various substrates, such as ruthenium,[4] copper,[5] and boron nitride,[6] have been produced successfully. A 0.26 eV band gap was opened for graphene sheets deposited on SiC substrate,[7] and was explained theoretically.[8,9] In practical application, graphene-based nano devices must be supported on substrates, which impact on the electronic properties of graphene.

Yet for meaningful applications, regularly patterned graphene nanoribbons (GNRs), instead of graphene sheets, should be more favorable in quasi-one-dimensional nano devices. GNRs were manufactured by chemical etching,[10] surface-assisted coupling of molecular precursors,[11] and unzipping of carbon nanotubes.[12] Moreover, GNRs on $SiO_2$, the most popular dielectric medium in integrated circuits,[13] have been applied into devices. For example, field effect of sub-10 nm GNRs on $SiO_2$ substrate has been explored,[14] and Bai's group produced 6-10 nm GNR-based field effect transistors (FETs) with $SiO_2$ serving as dielectric material.[15] In these systems, band gap is opened for GNRs. One issue is then aroused, what role of $SiO_2$ substrate is playing in affecting the electronic properties of GNRs. Actually, graphene sheets on $SiO_2$ are found by some calculations to have strong coupling with the substrate and become insulated when placed on O-terminated surfaces, whereas graphene sheets on Si-terminated surface maintain metallic properties due to weak interaction between C atoms and Si atoms.[13, 16] The same phenomenon may appear for GNRs. Also, electronic structures could be adjusted by controlling the edge states and widths of GNRs, as several calculations suggest.[17, 18] Yet, no first-principles calculations have explored the electronic properties of edge-functionalized GNRs on $SiO_2$ substrate.

---

† E-mail: miaoling@mail.hust.edu.cn

In this work, electronic properties of edge-functionalized ZGNRs on $SiO_2$ substrate were studied via first-principles calculations. Two kinds of models were discussed, H-ZGNRs directly laid on $SiO_2$ (0001) surfaces, and bivalent functional groups linking edges of ZGNRs and substrate. The former model exhibit distinct properties of H-ZGNRs due to different substrate-H-ZGNRs interactions. Bivalent functional groups (O, NH, $CH_2$, BH) decorating edges of ZGNRs act as the bridge to connect ZGNRs and the substrate, changing edge states and modifying the electronic properties of ZGNRs. These results may provide clues on how the substrate and edge states modulate GNRs properties.

## II. CALCULATION METHOD

First-principles calculations based on density functional theory (DFT) [19, 20] were carried out by SIESTA code,[21] which implements the linear combination of atomic orbitals (LCAO) method.[22] The double-$\zeta$ basis set was adopted to ensure a good computational convergence. Generalized gradient approximation (GGA) was used with Perdew–Burke–Ernzerhof (PBE) exchange-correlation functional.[23] Troullier–Martins scheme was employed for the norm-conserving pseudopotentials to represent the interaction between localized pseudoatomic orbitals and ionic cores.[24] The Brillouin zone sampling was performed using a set of $k$ points generated by the 1×4×1 Monkhorst-Pack grid.[25] The energy cutoff was 200 Ry, and atom positions were fully relaxed until the force on each atom was less than 0.05 eV/ Å.

ZGNRs were placed on different (0001) surfaces of α-quarts $SiO_2$ in hexagonal crystal system, with lattice constants of $a=b$=4.913 Å. $SiO_2$ slab was thicker than 7 Å to guarantee that it maintains bulk properties,[26] and the back side of the slab was passivated by hydrogen. The edges of ZGNRs were also passivated by hydrogen and decorated by various bivalent functional groups. A vacuum region ranging from 15 to 25 Å along $c$ direction was set up upon the model and the distance between edges of ZGNRs in adjacent periodic box along $a$ direction was about 10 Å.

Here we use n-ZGNRs (n-H-ZGNRs) to represent ZGNRs (H-ZGNRs) whose width is n. There are three different $SiO_2$ (0001) surfaces, two O-terminated surfaces (O1 surface and O2 surface) and a Si-terminated surface (Si surface) [FIG. 1(a)]. For calculation convenience, surface Si atoms that do not bond ZGNRs were passivated by hydrogen in all models.

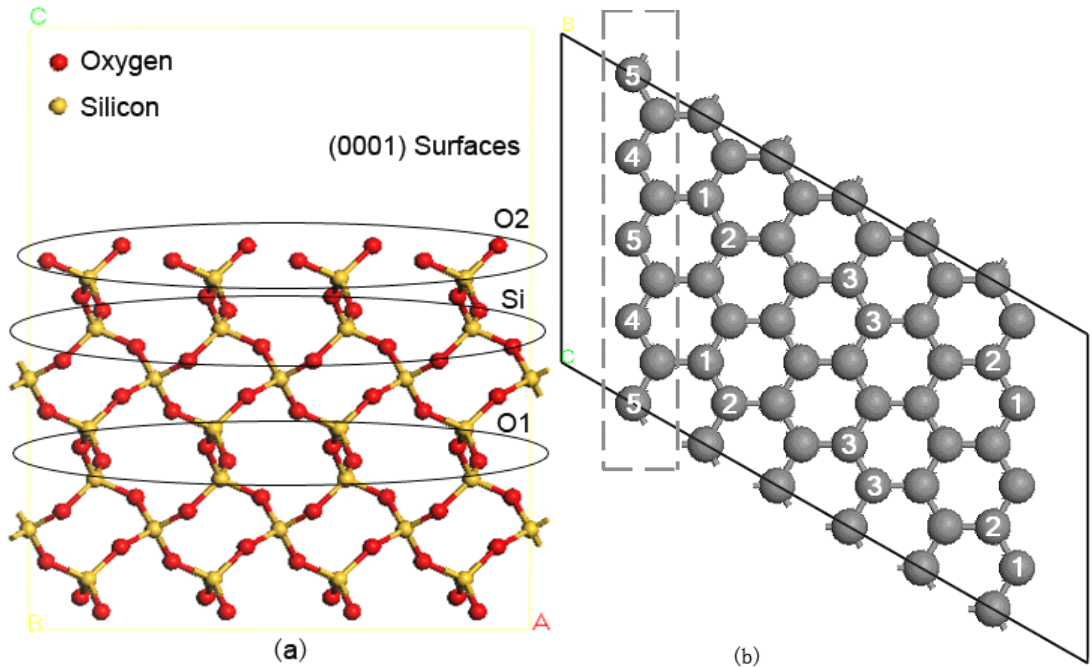

FIG. 1. (a) Three different $SiO_2$ (0001) surfaces: O1 surface, O2 surface and Si surface. The red and yellow balls represent Si and O atoms, respectively. (b) 6-ZGNRs and 5-ZGNRs(cutting off atoms in dotted rectangle). Atoms that may interact with substrate are labeled in numbers.

## III. RESULTS AND DISCUSSION

### A. H-ZGNRs placed on $SiO_2$ (0001) surfaces directly

We begin with the simplest structure, i.e. H-ZGNRs directly placed on $SiO_2$ O1 surfaces. The lattice constants of SiO$_2$ are almost as twice as that of graphene, hence only part of the atoms of ZGNRs [FIG. 1(b)] attach to the substrate directly. After geometrical optimization, No.1, 2, and 3 atoms of H-ZGNRs bond surface oxygen atoms of SiO$_2$ substrate and become sp$^3$ hybridization [FIG. 2(a)]. Structure analysis revealed that both surfaces of the substrate and the ZGNRs are distorted. The average of nearest C-O distances is 1.38 Å (less than 1.5 Å), indicating oxygen-carbon covalent bonding.

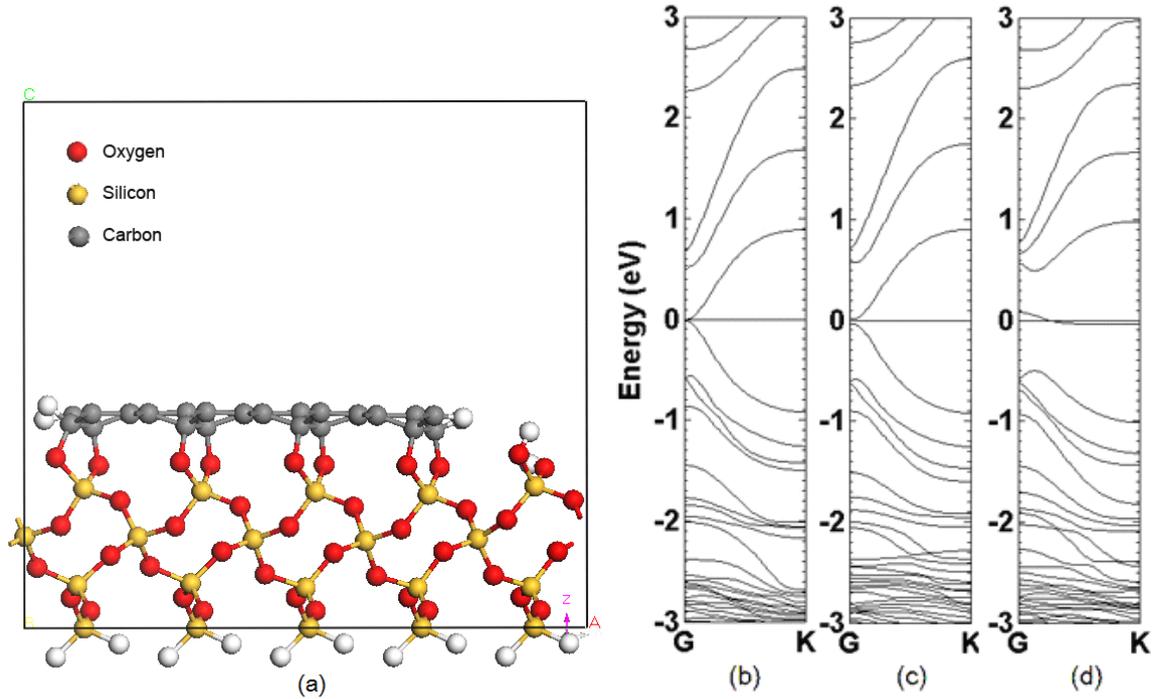

FIG. 2. (a) Structure for 7-H-ZGNRs on O1 surface. The red, yellow, gray, and white balls denote the O, Si, C, H atoms, respectively. Band structure of (b) 7-H-ZGNRs on O1 surface, (c) 7-H-ZGNRs on O2 surface and (d) 8-H-ZGNRs on O2 surface. Fermi levels are set to zero.

In the calculated band structure of 7-H-ZGNRs on O1 surface [FIG. 2(b)], no band gap is observed and bands crossed Fermi level at $\Gamma$ point, showing metallic properties. Free-standing H-ZGNRs were found to be metallic by some previous researches, yet recent studies believe that H-ZGNRs behave semiconductively owing to edge magnetization.[27] In our system, SiO$_2$ substrate pins the edges of H-ZGNRs and suppresses the effects, so H-ZGNRs remain metallic. Compare band structures of H-ZGNRs on O1 surfaces and that of free-standing ones, the only disparity lies in the shaded projected bands of SiO$_2$ substrate 3 eV below Fermi levels. High similarity indicates that properties of H-ZGNRs are hardly affected by the bonding to O1 surface and thus remain metallic. This metallic behavior is distinct from that of graphene sheet on O-terminated surface, which become insulators caused by the strong coupling of graphene to the substrate, according to some investigators.[13, 16] For further probe of the distinction, band structure of graphene sheet on O1 surface was calculated, which revealed obvious metallic property. In fact, $SiO_2$ (0001) surfaces contain two kinds of O-terminated surfaces and they interact differently to graphene, leading to dissimilar band structures. Only one surface, O2 surface, was discussed by previous researches. Binding energies between O1/O2 surfaces with H-ZGNRs were also calculated, showing more stable H-ZGNRs on O1 surface than O2 surface.

H-ZGNRs supported on O2 surface also make covalent bonds to surface. For odd-width H-ZGNRs, edge atoms lay right above the O atoms on surface, resulting in No.1,2,3 atoms relaxing toward O atoms and forming C-O bonds with bond distance 1.44 Å. Deformation induced by strong coupling between H-ZGNRs and O2 surface breaks chemically active $\pi$-orbital network among carbon atoms. Band structures of odd-width H-ZGNRs on O2 surface resemble that of substrate-free H-ZGNRs, exhibiting zero-gap feature. The influence of interaction, which splits $\pi$ bonding among carbon atoms and turns edge states into $sp^3$ hybridization, only introduces relatively small band gaps among narrow odd-width H-ZGNRs, as shown in FIG. 2(c). As width increases, the band gap decreases correspondingly [FIG. 3]. 7-H-ZGNRs and wider H-ZGNRs become metallic. Zero band gap of H-ZGNRs differs from insulating band gap of graphene sheets on O2 surface. For even-width H-ZGNRs, only one edge lies right above the O atoms and No. 1,2,3,4 atoms link to substrate [FIG. 1 (b)]. The band structure is presented in FIG. 2(d), in which a localized band around Fermi level appears. Apart from that, the degeneracy at the Dirac point is eliminated with the energy splitting of around 1.0 eV, which reduces when the width gets larger [FIG. 3]. Wider ZGNRs were calculated with hydroxyl substituting for substrate, for the convenience and efficiency of calculation. The substitution has little effect on the band structure of ZGNRs on $SiO_2$ substrate [FIG. 5(c)], and functions as an efficient method for investigating large graphene-substrate systems.

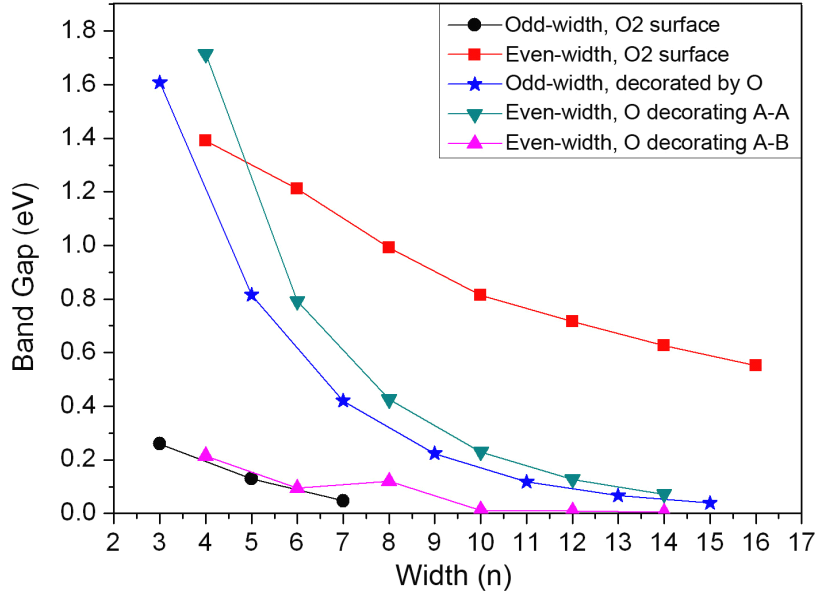

FIG. 3. The *Width-Band Gap* curves in structures that open band gap. The general trend is that with the increase of width of H-ZGNRs, band gap decreases.

Beside two O-terminated surfaces, H-ZGNRs on Si surface were also studied. Graphene sheets placed on Si-terminated surface are believed to have no interaction with Si atoms and preserve metallic properties.[13, 16] Likewise, H-ZGNRs pull themselves away from the substrate after geometric optimization and C atoms position above Si atoms at a distance of 2.48 Å from the substrate, indicating no covalent bonding. The linear bands of free-standing H-ZGNRs are generally maintained, with bands intersecting at Fermi levels.

### B. Bivalent functional groups linking edges of ZGNRs and substrate

The electronic properties of ZGNRs closely relate to edge states,[28, 29] thus could be modified by decorating edges with proper functional groups.[30, 31] Bivalent functional groups link active edges of ZGNRs and the substrate, and modify the electronic properties of ZGNRs. The case that O linking ZGNRs and substrate was mainly investigated, considering the uniform effect of O, NH, $CH_2$ and BH to the band structures of ZGNRs in our calculation. The lattice constants of $SiO_2$ double that of graphene, therefore only every other edge carbon atoms form covalent bonds with O to get connected to the substrate.

O decorating edges of odd-width H-ZGNRs is discussed first [Fig. 4(a)]. No.1 atoms on edges [FIG. 1(b)] bond to O by $sp^3$ hybridization. After geometry optimization, H-ZGRNs bends into arch with edge-substrate distance 2.36 Å and middle-substrate distance 3.27 Å. In band structures of 7-H-ZGNRs decorated by O [FIG. 4 (b)], a band gap of about 0.4 eV is opened. Sp$^3$ hybridization in edge atoms perturbs edge states and energy splitting is induced. Also, arched H-ZGNRs bring in stress, leading to band gaps. As GNRs get wider, their properties gradually approach graphene sheets', and edge states contribute less to determining the electronic properties. [28] Accordingly, band gap decreases as odd-width H-ZGNRs get wider [FIG. 3].

For even-width H-ZGNRs, it is more complicated since they have two fashions to get edge-modified by O [illustration in FIG. 4. (a)], edge atoms oppositely bond to O (A-A), or staggeredly bond to O (A-B). Different edge states are then brought in and distinct electronic properties are exhibited. Even-width H-ZGNRs' opposite A-A edge atoms bonding O resembles odd-width H-ZGNRs in electronic performance. As shown in FIG. 4 (c), band gap opened in 8-H-ZGNRs is approximately 0.4 eV. Edge states heavily change, with sp$^3$ hybridization of opposite edge atoms (A-A) co-affecting localized $\pi$ electrons in middle, resulting in a comparatively large band gap. Again, the band gap gets smaller with the increase of width. For staggered edge atoms (A-B) bonding to O, however, parallel to the band structures of free-standing H-ZGNRs, bands intersect at $X$ point of Fermi level and linear $E$-$k$ curve is generally maintained [FIG. 4 (d)].

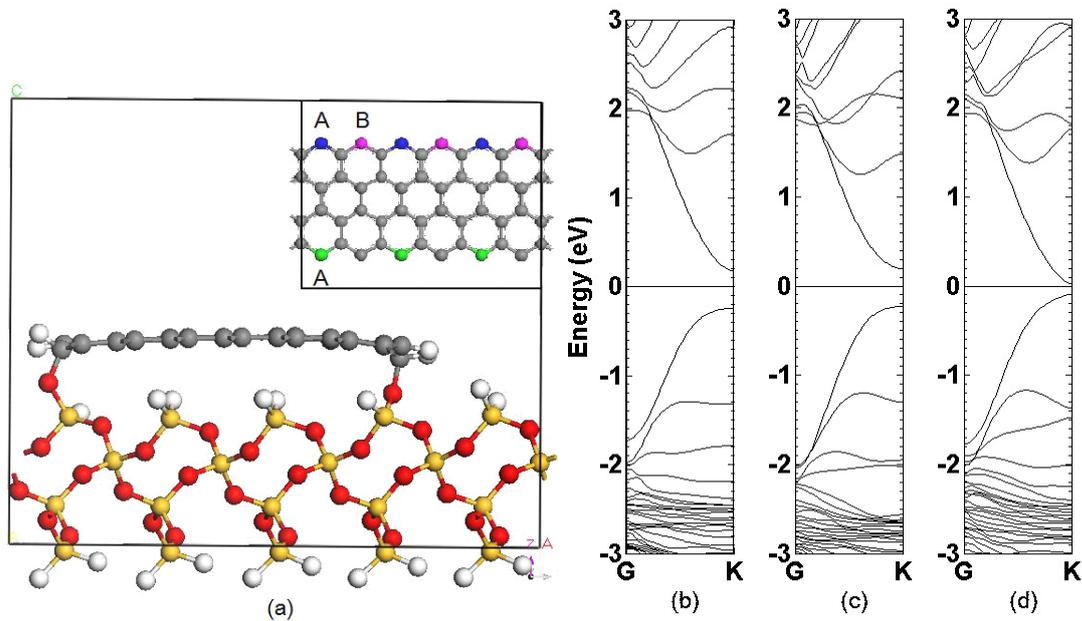

FIG. 4. (a) Structure of O linking edges of 7-H-ZGNRs and Si surface; the illustration shows two ways that even-width H-ZGNRs are decorated by bivalent functional groups: opposite A-A, or staggered A-B edge atoms make covalent bondings to O. The band structures of: (b) 7-H-ZGNRs, (c) A-A and (d) A-B edge atoms of 8-H-ZGNRs decorated by O.

Another manner of edge functionalization is to decorate edge atoms and near-edge atoms (No. 1 and 2

atoms in FIG. 1 (b)) by O respectively. The bonded edge and near-edge atoms are converted into $sp^3$ hybridization. Distorted lattice stems from the interaction, with edges approaching substrate to 2.28 Å and middle atoms leaving substrate to 3.36 Å. Typical metallic band structures include a band going through Fermi levels, and another band reaching Fermi levels [FIG. 5 (a)]. This situation is quite interesting. On one hand, H-ZGNRs are "pasted" to substrate stably; on the other hand, the strong coupling to the substrate has little effect to the characteristics of H-ZGNRs, even when edge states are heavily disturbed.

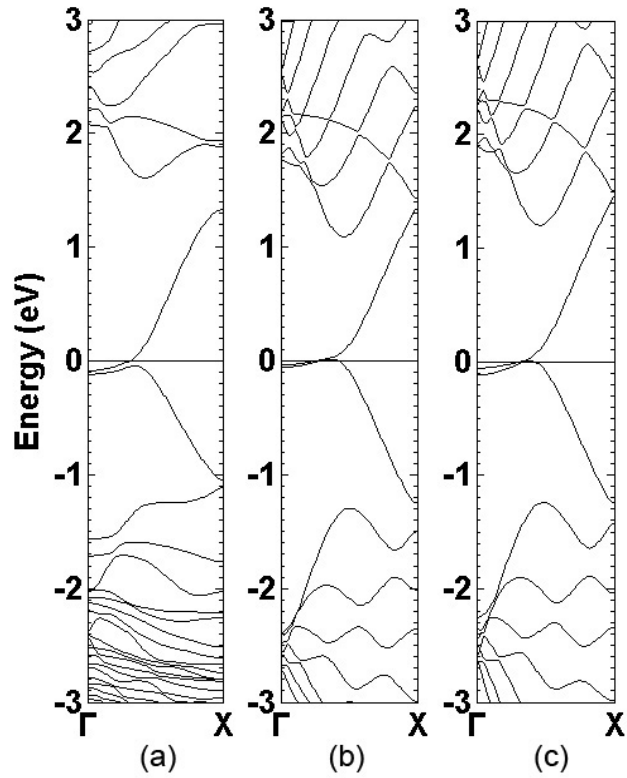

FIG. 5. Band structures of: (a) edge and near edge atoms (No. 1 and 2 atoms in FIG. 1 (b)) of 7-H-ZGNRs are modified by O respectivly to get connected to substrate. (b) O linking 7-ZGNRs and substrate, with all carbon atoms maintaining $sp^2$ hybridization. (c) Substuting substrate in situation (b) with hydroxy. (b) and (c) are generally the same, suggesting that the substitution has little effect to the electrnic properties of the substrate-edge-functionalized-ZGNRs system and could be used to investigate larger geometric structure.

All the structures discussed above have destroyed edge states, in which half edge atoms are $sp^3$ hybridized. To further understand whether edge states turning from $sp^2$ to $sp^3$ hybridization is one of the reasons for ZGNRs on $SiO_2$ to open a band gap, we studied another structure where all carbon atoms stay $sp^2$ hybridized. Bivalent functional groups are still used to link edges of ZGNRs and substrate, and only none-functionalized edge atoms are hydrogenated. ZGNRs arch on substrate, with edges 2.31 Å and middle atoms 3.51 Å away from substrate. sp$^2$ hybridization is maintained, and $\pi$ bonding among carbon atoms stays active. Electronic properties bear a resemblance to substrate-free ZGNRs and linear *E-k* curve intersects across Fermi levels [FIG. 5 (b)]. In summary, edge-functionalized ZGNRs with all carbon atoms sp$^2$ hybridized exhibit free-standing graphene electronic characteristics, and at the same time stand on SiO$_2$ with bivalent functional groups acting as a bridge.

## IV. CONCLUSIONS

In this work, electronic properties of edge-functionalized ZGNRs on SiO$_2$ (0001) surfaces were investigated by first-principles calculations. We found that H-ZGNRs have strong interaction with two kinds of SiO$_2$ (0001) O-terminated surfaces by the oxygen-carbon covalent bonds. Interestingly, the case H-ZGNRs on O1 surface was more stable, and remain metallic, quite different from previous researches on graphene sheets. Some structures such as, H-ZGNRs whose edge and near-edge atoms are functionalized by bivalent functional groups, and edge-modified ZGNRs with all atoms sp$^2$ hybridized, also stabilized on SiO$_2$ substrate and remain metallic. Meanwhile, some other structures such as, odd-width H-ZGNRs on O2 surface, and even-width H-ZGNRs with staggered edge atoms (A-B atoms in FIG. 4 (a)) bonding to bivalent functional groups, exhibited varied electronic properties depending on edge states and width of H-ZGNRs. Moreover, the former could be introduced a relatively large band gap of about 1.0 eV, resulting from edge states changing to sp$^3$ hybridization and distorted lattice by interaction with substrate. As the width of H-ZGNRs get wider, the structures have narrower electronic band gaps and display a transition from semiconductive to metallic characteristics. The ZGNRs-on-quartz system exhibits a mixture of metal and semiconductor behaviors, could be applied to multifarious graphene-based nano electronic devices as conducting wire and FETs, etc. The results show a feasibility to fabricate whole nano integrated circuits on a insulated SiO$_2$ substrate.

## ACKNOWLEDGEMENTS

This research work is supported by Innovative Foundation of Huazhong University of Science and Technology (Grant No. C2009Q007). Computational resources provided by Center of Computational Material Design and Measurement Simulation, Huazhong University of Science and Technology.